\documentclass[aps,prd,showpacs,nofootinbib,floats,floatfix,preprintnumbers,
groupedaddress,twocolumn]{revtex4}

\usepackage{bm}
\usepackage{latexsym, environ}
\usepackage{dcolumn, braket}
\usepackage{amsmath,amsfonts,amssymb}
\usepackage{graphicx,epsfig}
\usepackage{fancyhdr}
\usepackage{hyperref}
\usepackage{graphicx,epstopdf}
\hypersetup{
	colorlinks   = true, 
	urlcolor     = blue, 
	linkcolor    = blue, 
	citecolor   = red 
}

\def\r{\ref}
\def\no{\nonumber}
 
\def\o{\omega}

\def\r{\ref}
\def\no{\nonumber}
\def\a{\alpha}
\def\b{\beta}

\def\D{\Delta}

\def\z{\zeta}
\def\t{\tau}

\def\f{\frac}

\newcommand{\be}{\begin{equation}}
\newcommand{\ee}{\end{equation}}
\newcommand{\bes}{\begin{equation*}}
\newcommand{\ees}{\end{equation*}}

\def\l{\left}
\def\r{\right}

\NewEnviron{eqn}{
\begin{align}
\begin{split}
  \BODY
\end{split}
\end{align}
}

\NewEnviron{eqn*}{
\begin{align*}
\begin{split}
  \BODY
\end{split}
\end{align*}
}

\newcommand{\ms}{\mathfrak{s}}

\begin{document}
\title{Fluctuation-dissipation in accelerated frames}
\author{Ananya Adhikari\footnote {\color{blue} a.adhikari003@gmail.com}}
\author{Krishnakanta Bhattacharya\footnote {\color{blue} krishnakanta@iitg.ernet.in}}
\author{Chandramouli Chowdhury\footnote {\color{blue} chandramouli@iitg.ernet.in}}
\author{Bibhas Ranjan Majhi\footnote {\color{blue} bibhas.majhi@iitg.ernet.in}}

\affiliation{Department of Physics, Indian Institute of Technology Guwahati, Guwahati 781039, Assam, India}

\date{\today}

\begin{abstract}
 An uniformly accelerated (Rindler) observer will detect particles in the Minkowski vacuum, known as Unruh effect. The spectrum is thermal and the temperature is given by that of the Killing horizon, which is proportional to the acceleration. 
Considering these particles are kept in a thermal bath with this temperature, we find that the correlation function of the random force due to radiation acting on the particles as measured by the accelerated frame, shows the fluctuation-dissipation relation. 
It is observed that the correlations, in both ($1+1$) spacetime and ($1+3$) dimensional spacetimes, are of Brownian type. We discuss the implications of this new observation at the end. 
\end{abstract}
\maketitle
\noindent
{\bf I. Introduction:}
Instances of Brownian motion, i.e., the non-equilibrium random motion of a particle immersed in a bath, have been known for quite long time. Iconic traditional examples include the motion of a macro-particle (such as a pollen grain) immersed in water or the motion of suspended dust particles in air. The non-equilibrium statistical behavior of such particles executing Brownian motion is governed by the Langevin equation and the fluctuation-dissipation theorem. The Langevin equation shows that the driving force behind the motion of the immersed particle can be decomposed into three parts - (a) the external force exerted by external agents that the particle is subjected to, (b) the slowly-varying averaged force that tends to drive the system to an equilibrium state, and (c) the rapidly fluctuating random force. The fluctuation-dissipation theorem expresses the dissipative coefficient of the system with the correlation function of the fluctuating random component of the force (see Refs. \cite{reif, kubo:1985}).

In Rindler spacetime, i.e., an uniformly accelerated observer in Minkowski spacetime, it is known that the observer perceives a thermal bath of uniform temperature emerging from the Minkowski vacuum \cite{Unruh:1976db}. The emitted spectra from the Killing Rindler horizon is thermal in nature and its temperature is proportional to the value of the acceleration. This observation has significant impact in gravitational theories. As equivalence theorem implies that an accelerated frame can mimic gravity, the same can happen in gravitational theories as well. This has been supported by the discovery of Hawking radiation \cite{Hawking:1974rv}. On the other hand, {\it locally} the metric is given by a null one which, in general, is not a solution of Einstein's equations. Under certain assumptions, this is taken to be the Rindler metric. Since one can associate temperature and entropy on the horizon, it has been found that the Einstein's equation can be obtained from the first law of thermodynamics \cite{Jacobson:1995ab}, revealing the emergent nature of gravity (see \cite{Padmanabhan:2010xe}, for a review on this topic). 
Moreover, the metric in the near horizon region of a non-extremal black-hole can be represented by the Rindler form. It is, therefore, quite obvious that such a metric plays an important role in exploring the nature of gravity.  Only the thing one has to remember is that the accelerated observer plays the role of static observer in the case of true gravity.

In this situation, the emitted particles from the horizon can be considered as the system of particles immersed  in a thermal bath with temperature determined by the surface gravity that of the horizon. Now the question is: What is the behavior of these particles? This can be, in principle, obtained by finding the force on each one and then solving these force equations. As we have huge number of particles in the bath, the statistical calculation will be fruitful. In this paper we address this issue in details. The idea is to calculate the correlation function corresponding to the random force exerted on particles and see its nature.

Here we consider the system of Unruh radiating particles and a separate analysis is done for the underlying background being ($1+1$) as well as ($1+3$) dimensional. We treat the radiation bath as a simple massless scalar field. 
With this simple construction, the correlation function of the random force in $(1+1)$- and ($1+3$) dimensional Rindler frame (which is the proper frame of the accelerated observer) can be determined.
Important point to note is that-- the scalar field themselves are in the Minkowski vacuum and, therefore, while calculating the correlation functions of the quantities defined in the Rindler frame, we shall take the expectation value with respect to the Minkowski vacuum.  
It will be then observed that the correlation function of the random force exhibits the fluctuation-dissipation theorem. 
Both, in the ($1+1$) and ($1+3$)-dimensional cases, the correlation functions 
depend only on the difference of the two time arguments, when expressed in 
terms of the proper time of the detector. We shall see that this fact plays an important role in deducing that the motion of the particle in this case is Brownian, as 
it satisfies the fluctuation-dissipation theorem. 
Finally, we calculate the coefficient of the mean dissipative force by using the 
standard relation in non-equilibrium statistical mechanics.
It is found to be dependent on both, the dimension and the direction. In the case of $(1+1)$-dimension we obtain that it is proportional to the fourth power of the temperature. 
On the other hand, in the $(1+3)$-dimensional case, we see that the longitudinal and
the transverse component of the coefficient of the mean dissipative force is 
proportional to the sixth and the eight power of the temperature respectively. 
This is in contrast to the  situation that was observed for a moving mirror 
in Minkowski spacetime when the velocity of the mirror is small; i.e. in 
the non-relativistic limit \cite{Stargen:2016euf, ford}. 
However, this does not mimic gravity. There is another earlier attempt \cite{Kolekar:2012sf} to explore the issue; but it was studied in a Rindler-boosted frame in ($1+3$) dimensions and the approach is completely different from the present one (also see \cite{Caceres:2010rm, Iso:2011gb, Rigopoulos:2016oko} for different situations). 

The present observation is completely new and will have wide range of 
implications in the theory of gravity. It shows that the motion of the radiating particles is very much random and is accompanied by the fluctuation-dissipation theorem. As a result, one would think that the dynamics of this is governed by non-equilibrium statistics. If this is so, the nature of emitted particles, as a whole, can be studied with the tools of existing formalisms. Moreover, as the near horizon geometry of a non-extremal black hole is Rindler in nature, the obtained result in principle explores Hawking radiating particles. On top of that, it must be noted 
that this whole argument is observer dependent, as only the accelerated 
frame (static observer for gravity case) feels it. So the statistics will have 
an observer dependence which is a clear distinction over the usual existing formalism. This fact is similar to the observer dependence of the thermodynamic quantities in the context of gravity (see \cite{Majhi:2012tf,Majhi:2013jpk,Majhi:2017fua}, for more on this).    

Let us now proceed to the main analysis.
We have worked with the units $c = \hslash = k_B = 1$. Angular brackets denote expectation values at finite temperature. The metric signature is taken to be ($-,+$) in 2-dimensions and ($-, +, +, +$) in 4-dimensions.
\vskip 2mm
\noindent
{\bf II. Correlation of random force:}
Here we shall find the random force acting on the radiating particles, as seen by the accelerated frame. Also the correlation function of this force will be evaluated which will be the main inputs of our claim. Both, the ($1+1$) and ($1+3$) cases will be discussed separately.

\vskip 1mm
\noindent
\underline{\it{Case I: $(1+1)$ spacetime}}: 
 To start our analysis, we need a set of coordinates which is useful to describe an accelerating observer. In GR, a particle with a proper acceleration, moving in  Minkowski spacetime, is described conveniently in terms of the Rindler coordinates, which are given as follows:
\begin{equation}
X = \f{1}{a} e^{ax} \cosh(at); \ \ T = \f{1}{a} e^{ax} \sinh(at)~. \label{2DRIND}
\end{equation}
Here $a$ denotes the acceleration of the moving observer. Also, $(T, X)$ denotes the Minkowski coordinates, and $(t, x)$ denotes the Rindler coordinates. For convenience, we shall introduce the null coordinates: $U = T - X$ and $V = T + X$ with $u = t - x$ and $v = t + x$. Then from \eqref{2DRIND} one obtains $U = -\f{1}{a} e^{-a u}$ and $V = \f{1}{a} e^{a v}$. The metric in the null coordinates is given as
\begin{equation}
\,ds^2_{(1+1)} = - \,dU \,dV = - e^{a (v - u)} \,du \,dv~. \label{2DMETRIC}
\end{equation}
The Rindler horizon is located at $T=X$.

 Now, there is a momentum associated with the scalar field radiation, that the detector is immersed in. This momentum, in Rindler spacetime, is given by 
\begin{equation}
p = l_x \int d{x} \  T^{01} \delta(x - x_D) = l_x T^{01}(\tau) \label{2DMOM}
\end{equation}
Here, $\delta(x - x_D)$ indicates that the detector we are considering is static with respect to the Rindler observer, with the position of the detector being represented by $x_D$ and its proper time being denoted by $\tau$. In other words, the Rindler observer is the detector itself. 
Thus, the argument of time in the stress tensor becomes the same as the detector's proper time $\t$. 
Here, $l_x$ is just an arbitrary length scale, introduced to maintain correct dimension. In principle this can be taken as any length scale appearing in the present system; i.e. it may be the size of the detector. Due to this momentum, the moving observer experience a force per unit length $(l_x)$, $F$, which is given as,
\begin{equation}
F = \frac{d T^{01}}{d\tau} \label{2DFORCE}~.
\end{equation}
Here, the Rindler frame is the proper frame of the moving observer, implying that there is no spatial displacement (i.e. $x=0$) of the observer with respect to the mentioned frame. Therefore, it implies the proper time is given as $\t=u=v$.
The above force can be interpreted as the radiation force experienced by the detector as it will see creation of particles in the Minskowski vacuum with a temperature given by $T=a/2\pi$.

Let us now concentrate on the component of the energy-momentum tensor appearing in the above equation \eqref{2DFORCE}. From the rules of the coordinate transformations, one can straightforwardly obtain
\begin{equation}
T^{01} \equiv T^{t x} = \f{1}{4} \big( T^{vv} - T^{uu} \big)=\frac{1}{4}(T_{uu}-T_{vv})~, \label{T012D}
\end{equation}
in the Rindler proper time.
Note that the above consists of two parts -- the term $T_{uu}$ corresponds to the outgoing modes, which gives the radiation flux (Unruh radiation) whereas, the term $T_{vv}$ is associated to the ingoing modes. Since we are interested in Unruh radiation, only the $T_{uu}$ part is relevant for our discussion. Hence, we concentrate only on  
$T^{t x}_{out} = \f{1}{4} T_{uu}$.

Note that with respect to the Rindler frame, the system consists of Unruh particles and the detector (which is the Rindler frame itself), immersed in a thermal bath of temperature given by the Unruh expression. These particles are carrying a momentum, defined in (\ref{2DMOM}), as measured by the moving frame. This momentum, that we shall consider later, corresponds to the energy-mometum tensor which leads to Unruh radiation. 
The corresponding force, therefore, we call as radiation force. This is motivated by the ideas of radiation pressure exerted on an object by a beam of particles, which in this case is a thermal beam. Here the concept of particles, and the idea of a thermal beam is not ill-defined as the Unruh-Fulling effect arises. And in this present analysis we are interested in the contribution of only such a radiation force. The possible source of this force may be due to the ``hard-sphere'' type interaction between the fields and the detector. Motivated by this we are only concentrating on the energy-momentum tensor of the resulting radiation which allows us to quantify the force. 

From the above discussion, we can now quantify the random fluctuating force, $R(\t)$, acting upon the particle is given by
\begin{equation}
R(\t) = F(\t) - \langle F(\t)\rangle~,
\label{R2D}
\end{equation}
where the last term is the vacuum average of the force. In our present case, the vacumm is the Minkowski one which is denoted by $|0\rangle$.
Since, $\langle F(\t)\rangle \equiv \f{1}{4} \frac{d}{d\t}\langle T_{uu}\rangle$ it gives, $\langle F(\t)\rangle= 0$, implying $R(\t)=F(\t)$ and, therefore, the correlation function of the random force, $\langle 0| R(\t)R(\t')|0\rangle$, evaluated in the inertial frame is given by
\begin{align}
\langle 0|R(\t)R(\t')|0\rangle=\f{1}{16}\frac{d^2}{d\t d\t'}\langle 0| T_{uu}(\t)T_{u'u'}(\t')|0\rangle~. \label{R2DCOR}
\end{align}
In the above equation, $|0\rangle$ represents the Minkowski vacuum.
To evaluate the above correlation function, one needs to calculate $\langle 0|T_{uu}(\t)T_{u'u'}(\t')|0\rangle$ explicitly. It can be obtained explicitly from the Schwinger function, the most general form of which is given by \cite{francesco} as
\begin{equation}
\begin{split}
&S_{m n r s}(\textbf{\textit{x}}_1,\textbf{\textit{x}}_2) = \left< T_{m n}[\textbf{\textit{x}}_1]T_{r s}[\textbf{\textit{x}}_2]\right> \\ & = \frac{A}{\left(\tilde{\textbf{\textit{x}}}^2\right)^4}[ \left( 3g_{m n}g_{r s}-g_{m r}g_{n s}-g_{m s}g_{n r} \right)\left(\tilde{\textbf{\textit{x}}}^2\right)^2 \\ &- 4 \tilde{\textbf{\textit{x}}}^2 \left( g_{m n} {\tilde{x}_r} \tilde{x}_s + g_{r s} \tilde{x}_m \tilde{x}_n \right)+8 \tilde{x}_m \tilde{x}_n \tilde{x}_r \tilde{x}_s ]~,
\label{Schwinger}
\end{split}
\end{equation}
where $A$ is an arbitrary constant, related to the central charge $C$ of the particular fields by $A=C/4\pi^2$, $\tilde{x}^a = x^a_1 - x^a_2$ and $\tilde{{\textbf{\textit{x}}}}^2= -(x^0_1-x^0_2)^2+(x^1_1-x^1_2)^2$. Since for the present case we have considered only massless scalar fields, its value is given by $A=1/4\pi^2$ as $C=1$. Using $T_{uu}=e^{2 a u} T_{UU}$, one can obtain 
\begin{equation}
\langle 0|T_{uu}(\t)T_{u'u'}(\t')|0\rangle = \f{a^4}{256 \ \pi^4} \f{1}{\sinh^4(\f{1}{2} a \D\t)} ~,
\label{TuuTuu}
\end{equation}
where $\D\t=\t'-\t$. We shall use this later for our main purpose.

\vskip 1mm
\noindent
\underline{\it{Case II: $(1+3)$ spacetime}}:
The above analysis in (1+3) dimensions is expected to be a bit different as, in this case, the observer moves along one direction and, there are two transverse directions.
If we consider the accelerated observer is moving along the $X$-direction, then it just follows the transformation rules mentioned in \eqref{2DRIND}, and the Rindler metric in the null coordinate can be written as 
\begin{align}
ds^2_{(1+3)}=-dUdV+dY^2+dZ^2=- e^{a (v - u)} \,du \,dv
\no 
\\
+dy^2+dz^2~.
\end{align}
Following the similar approach of the $2$-D spacetime, we can again define the Force per unit volume (experienced by the moving observer) here in a similar fashion. Since we are performing the calculation in the detector's frame of reference, we must have similar delta functions in this case as in eq.\eqref{2DMOM}, which picks out the detector's position. 
This will naturally introduce three length scales, along $x$, $y$ and $z$ directions, and thus, we consider the force per unit volume, along a particular direction which will be given as
\be
F^{\a} = \frac{dT^{0 \a}}{d\t}~.
\ee
Here, $\a \in \{ x, y, z \}$ denotes all the spatial indices. 
It is known that the value of $\langle T^{0\a}\rangle$ is a constant ($= 0$ when renormalized), which implies $\langle\frac{dT^{0\a}}{d\t} \rangle \equiv \frac{d}{d\t} \langle T^{0 \a}\rangle = 0$.
Hence, the random fluctuating force, defined similar to the Eq. \eqref{R2D}, turns out in $4$-Dimensions as $R^{\a}(\t) = F^{\a}(\t)$. Thus, the correlation function of the random force $\langle0| R^{\a}(\t) \ R^{\a}(\t')|0\rangle$ is given as follows
\be
\langle 0| R^{\a}(\t) \ R^{\a}(\t')|0\rangle= \frac{d^2}{d\t d\t'} \langle 0| T^{0\a}(\t) T^{0\a}(\t')|0\rangle \label{RES4D}
\ee 
Again, we need to calculate $\langle 0| T^{0\a}(\t) T^{0\a}(\t')|0\rangle$ explicitly for all values of $\a$. Following the arguments of \cite{Padmanabhan:1987rq}, we obtain $ \langle T^{ac}(x_1) T^{bd}(x_2)\rangle\equiv G^{abcd}(x_1 - x_2) $, where
\be
G^{abcd}(x) = \l(\frac{8 W^2}{s^8}\r) \bigg[A^{abcd}(x) + s^4 C^{abcd}(x) - s^2 B^{abcd}(x)\bigg]~, \label{GABCD}
\ee
with, $W = \f{1}{4\pi^2 s^2}$ and $s^2 = x_i x^i$. The values for $A^{inkm}, B^{inkm}$ and $C^{inkm}$ are given below,
\begin{align*}
\begin{split}
A^{inkm} =& \ 32\  x^i x^k x^m x^n \\
C^{inkm} =& \ g^{in} g^{km} + g^{kn} g^{im} + 4 g^{ik} g^{nm} \\
B^{inkm} =& \ 4 \ \big[g^{in} x^k x^m + g^{km} x^i x^n + g^{kn} x^i x^m \\
&\ \ + g^{im} x^k x^n + 2 g^{nm} x^i x^k + 2 g^{ik} x^n x^m \big]~.
\end{split}
\end{align*}
We shall see that the correlation function of the random fluctuating force depends on the direction of the acceleration of the moving observer. Firstly, let us calculate the correlation of the energy-momentum tensors along the acceleration (which in our case is the $X$-direction) and, thereafter, the same analysis follows along the transverse ($Y$ and $Z$) directions.

\vskip 1mm
{\it{X-Direction}:}
Along the direction of acceleration of the observer (longitudinal), the term which appears in the correlation function of \eqref{RES4D} is $\langle 0| T^{01}(x) T^{01}(x')|0\rangle$, which can be expressed in the Minkowski coordinates as $T^{01} = \f{1}{4} \ \l( T^{vv} - T^{uu} \r)$. Again, we are only interested in the outgoing modes, which are given by $T^{vv}$. Therefore we have $T^{01}_{out} \equiv \f{1}{4} T^{vv}$. Thus,
\be
\langle 0| T^{01}(x) T^{01}(x')|0\rangle = \f{e^{-2a(v+v')}}{16} \langle T^{VV}(u,v) T^{VV}(u',v')\rangle
\ee
Using the relation \eqref{GABCD}, we finally obtain
\be
\langle T^{01}(\t) T^{01}(\t')\rangle = \f{a^8}{2^8\pi^4} \ \f{1}{\ms^8}
\label{TxTx}
\ee
Here we have used the shorthand notation $\ms = \sinh\l[ \f{a}{2}(\t - \t') \r] $.

\vskip 1mm
{\it Y/Z-Direction}:
Along the transverse direction, one can follow an exactly similar procedure as the longitudinal direction. The stress tensor as calculated in Rindler frame $(T^{ty})$, when expressed in terms of the tensor is Minkowski space, takes the following form,
\be
T^{ty} = \f{1}{a(T^2 - X^2)} \big[ T T^{XY} - X T^{TY} \big]~.
\ee
Thus using eq.\eqref{GABCD} we end up with, 
\be
\braket{T^{t\z}(\t) T^{t \z}(\t')} = -\f{a^8}{2^9 \pi^4 \ms^8} \big[ 3 + 2 \ms^2]
\ee
where, $\z \in (y, z)$. 
\vskip 2mm
\noindent
{\bf III. Fluctuation-Dissipation:}
In the earlier section, we have obtained the correlation function of the random force in (1+1) dimensions as well as (1+3) dimensions. In this section, we shall investigate them more closely. 
As it will be shown below, we obtain that the fluctuation-dissipation theorem is satisfied in both the $(1+1)$-dimensional case, and also the $(1+3)$-dimensional case. The $(1+1)$-dimensional case is pretty straightforward. 
In the $(1+3)$-dimensional case, one has to break the calculation into two parts- one, {\it the longitudinal component}, along the the direction of acceleration and two, {\it the transverse component(s)}, perpendicular to the direction of acceleration.
This implies that the particle subjected to such a random force will execute Brownian motion, in both the cases.  
The whole analysis is given as follows. 

\vskip 1mm
\noindent
{\underline{\it Case I: $(1+1)$ spacetime}}:
Calculation of the correlation function of the random force is now very straightforward. Using \eqref{R2DCOR} and \eqref{TuuTuu} we obtain
\begin{equation}
\braket{ 0| R(\t) R(\t')|0} = -\f{a^6}{2^{12}\pi^4} \bigg[\f{3}{\ms^6} + \f{4}{\ms^4}\bigg]~.
\label{1D-res}
\end{equation}
Note that expression above only depends on the difference of the proper times ($\Delta\tau$). Therefore, we can make a Fourier transformation of the correlation function in terms of this variable to check whether the fluctuation-dissipation theorem holds.

Let us define the correlation function $K(s)$ of any arbitrary function $F(t)$ in the time domain is given by 
$K(s)= \left< F(t) F(t+s) \right>$. Note that we are assuming that such a function is time-translationally invariant, which is consistent with the idea of Poincar\'e invariance of the Minkowski vacuum. 
Using this, we can define he symmetric and and anti-symmetric correlation functions as follows, 
\begin{equation}
\begin{split}
K^+(s) &= \frac{1}{2}[ \left< F(t_0) F(t_0 + s) \right> + \left< F(t_0 + s) F(t_0) \right> ] \\ &= \frac{1}{2}[K(s)+K(-s)]~,
\end{split}
\label{eq:corr+} 
\end{equation}
and 
\begin{equation}
\begin{split}
K^-(s) &= \frac{1}{2}[ \left< F(t_0) F(t_0 + s) \right> - \left< F(t_0 + s) F(t_0) \right> ] \\ &= \frac{1}{2}[K(s)-K(-s)]~.
\end{split}
\label{eq:corr-}
\end{equation}
In the case of Brownian motion, it has been observed that the Fourier modes, $\tilde{K}^+(\o)$ and $\tilde{K}^-(\o)$, of these in the frequency domain are related by \cite{ Stargen:2016euf, kubo}
\begin{equation}
\tilde{K}^+(\omega) = \coth\left(\frac{\beta\omega}{2} \right) \tilde{K}^-(\omega)~,
\label{FDTFREQ}
\end{equation}
where, $\b = \f{2 \pi}{a}$ is the inverse of the temperature at which the particles in the thermal bath is immersed in. This is known as fluctuation-dissipation theorem. 
The above celebrated relation, eq.\eqref{FDTFREQ}, is very important in non-equilibrium statistics and is the signature of the Brownian motion.
Here we shall investigate below if the earlier obtained correlation functions satisfy the same theorem.

For the present case, defining the Fourier Transformed version of the correlation function in our case as
\begin{equation}
K(\o) = \int \,d{\t} e^{ i \o \t} \braket{R(0) R(\t)}~,
\end{equation}
we obtain, 
\begin{equation}
K(\o) = \frac{f(\o)}{1-e^{-\b\o}}~,
\label{1D-fourier}
\end{equation}
where $f(\o)=\frac{a^4}{15\cdot2^{8}\pi^3}\l(\frac{3\o^5}{a^4}-\frac{\o^3}{a^2}-8\o\r)$.
Defining the symmetric and the anti-symmetric combinations as $\tilde{K}^+(\o)$ and $\tilde{K}^-(\o)$, as motivated from eq.\eqref{eq:corr+} and eq.\eqref{eq:corr-}, we find that their ratio comes to be $\coth\l(\b\o/2 \r)$, as in eq.\eqref{FDTFREQ}.
Therefore, the particles moving in the thermal bath in (1+1) dimensions shows fluctuation-dissipation relation and, the corresponding fluctuation is Brownian. 

Note that the Fourier transform of a function constructed out of the summation of terms only containing an even power of $\ms$ in the denominator (eg: eq.\eqref{1D-res}), shall give rise to an odd function of frequency in Fourier space, along with the Boltzman factor (eg: $f(\o)$ in eq.\eqref{1D-fourier}). This feature will also be observed in $(1+3)$-dimensions as well. 

Using the value of $\tilde{K}(\o)$, one can obtain the coefficient of the mean dissipative force acting on the particle. It is given by, 
\be
\a = \b \lim_{\b \to 0} \f{\tilde{K}(\o)}{2} \label{coefficient}~.
\ee
From eq. \eqref{1D-fourier} and \eqref{coefficient} it follows that the coefficient of mean dissipative force goes as $\a = -4\pi/(60\b^4)~.$

{\it\underline{ Case II: $(1+3)$ spacetime}}:
{\it X-Direction}:
From eq.\eqref{TxTx} we obtain the two point correlator of the response function along $X$ direction to be, 
\be
\langle R^{x}(\t) R^{x}(\t')\rangle = -\f{a^{10}}{2^7\pi^4} \bigg[ \f{9}{\ms^{10}} + \f{8}{\ms^8}  \bigg] \label{RES1}
\ee
This expression again is a function of the difference of the arguments of the two correlators. This enables one to calculate the Fourier transformation of the same, as a function of a single frequency, 
\be
K_x(\o) = \f{f_x(\o)}{1 - e^{-\b \o}}~. 
\label{X-fourier}
\ee
Here, $f_x(\o) = \f{2a^8}{7!\pi^3} \l( \f{36\o}{a^2} + \f{49\o^3}{a^4} + \f{14\o^5}{a^6}  +  \f{\o^7}{a^8}\r)$, which implies that the ratio of the symmetric construct of $K_x(\o)$ to that of the anti-symmetric construct goes as a $\coth(\b\o/2)$.
This is exactly of the form, eq.\eqref{FDTFREQ}, thus establishing the fluctuation-dissipation theorem along the longitudinal direction. 

The longitudinal coefficient of the mean-dissipative force, as evident from eq. \eqref{coefficient} and \eqref{X-fourier}, goes as $\a = 16\pi^3/(35\b^6)~.$

\vskip 1mm
{\it Y/Z-Direction}:
Proceeding in the exactly same manner as the $X$ direction, we can obtain the two-point correlator along the transverse directions, 
\be
\braket{R^{\z}(\t) R^{\z}(\t')} = \f{a^{10}}{2^9 \pi^4} \l[ \f{54}{\ms^{10}} + \f{69}{\ms^8} + \f{18}{\ms^6} \r] \label{RES2}
\ee
Here, $\z \in (y, z)$.  Now, following the calculation in the previous section, we obtain, 
\be
K_{\z}(\o) = \f{f_{\z}(\o)}{1 - e^{-\b\o}}~.
\label{Y-fourier}
\ee
Here, $f_{\z}(\o) = \f{a^{10}}{5!(4\pi^3)}\l(\frac{\omega ^9}{56 a^{10}}-\frac{31 \omega ^7}{28 a^8}-\frac{\omega ^5}{8 a^6}+\frac{169 \omega ^3}{7 a^4}+\frac{162 \omega }{7 a^2}\r)$. 
Thus, taking the ratio of the symmetric and the anti-symmetric constructs of the same, we obtain exactly the form of eq.\eqref{FDTFREQ}, thereby establishing the fluctuation-dissipation theorem along the transverse directions. 
The transverse coefficient of the mean-dissipative force as evident from eq. \eqref{coefficient} and \eqref{Y-fourier} is proportional to $216\pi^5/(35\b^8)~.$

Note that the features in parallel and transverse directions are different. Therefore, there exists an anisotropy in the correlation function of the random force.
A similar feature was also obtained earlier \cite{Kolekar:2012sf} in a slightly different analysis.
 

\vskip 2mm
\noindent
{\bf IV. Discussions and outlook}:
In this paper, we have studied the statistical behaviour of the particles (perceived from the uniformly accelerated Rindler frame), immersed in the constant temperature thermal bath which emitted from the Killing Rindler horizon. The temperature is taken to be equal to the horizon as the particles are in thermal equilibrium with this. 
It has been observed that the correlation function for the random force acting on the particles exhibits fluctuation-dissipation theorem. Both in 2-D and 4-D, the correlations are purely Brownian type as they satisfy the fluctuation-dissipation theorem.

The result implies that the emitted particles can behave similar to non-equilibrium statistics. Since the Rindler form can be taken as the near horizon geometry of a black hole and also the {\it locally} null surface, the observation has the following implication in gravity. The properties of the Hawking radiated particles can be studied in this direction. Moreover, a point to be noted is that the non-triviality occurs only for Rindler observer and, hence, the phenomenon is observer dependent. 

In this present analysis, the particles are taken to be scalar and the separate analysis has been made for ($1+1$)- and ($1+3$) dimensional spacetime. In principle, one can consider any type of particles. 
In 2-D we can expect that the final conclusion is same for all of them as the form of the Schwinger function (\ref{Schwinger}) does not change, except the value of the constant $A$ varies which is related to the central charge of the specific theory by the relation $A=C/4\pi^2$ with $C$ is the central charge (for example, see section $5.4$ of \cite{francesco}). But that has to be thoroughly studied. 
There are some issues needed to be investigated; like studying the similar problem for black hole spacetime, to get more insight. 
We can expect the similar results for those cases as well; because, it is well known that the near horizon geometry of a black hole is effectively ($1+1$) dimensions \cite{Robinson:2005pd, Iso:2006ut,Majhi:2011yi} and which at the lowest order is in Rindler form. In addition, it would be interesting to look for other possible observers which also can predict similar situation.

Finally, finding the Langevin equation in this case will be very crucial. Of course, one can guess the possible structure by using the coefficient for the mean dissipative force (\ref{coefficient}); but that needs deeper investigation to obtain a concrete form. 
Roughly speaking, for that purpose, one needs to incorporate all the forces arising in the system. 
There are a few works that can be mentioned which may help to write the equation.  Considering the test particle as a quantum harmonic oscillator and taking its coupling with the Unruh particles (since it is exactly solvable), people finds a Langevin type of equation (see \cite{Unruh:1989dd}, also see the subsequent papers \cite{Raine:1991kc, Hinterleitner:1992hp, Kim:1997fi, Kim:1998ne, Massar:2005vg, Massar:2006ev}). 
Similarly the same can be done here as well. For that one may consider the specific interaction term between the detector and the scalar fields in the action for the system. 
The most popular one is the monopole interaction of the detector with the scalar field (interaction Lagrangian is $\mathcal{L}_{I}= \alpha \mu \phi$, where the detector with its monopole moment $\mu$ couples to the scalar field $\phi$ directly and $\alpha$ is the coupling of the interaction).  
Here we found that the random force exerted by the radiation on the test particle shows fluctuation-dissipation theorem. Of course, one may include the interaction term. 
Then, depending on the coupling, the system might feel a force governed by different field operator, and it may not be the same derived in (2). In particular, it may depend on a coupling constant charactering the strength of the interaction (see e.g. \cite{Unruh:1989dd}). 
This would not affect our result significantly, since the vacuum state of a free field theory is nearly Gaussian, and higher order correlation functions are related to the two-point function by Wick's theorem. Thus, for the examination of the Brownian property, our method of calculating the two point correlation function of the random force is quite admissible.
The research in these directions are going on. Hope soon we shall be able to report more on this. 

\vskip 4mm
{\section*{Acknowledgments}}
\noindent  
The research of one of the authors (BRM) is supported by a START-UP RESEARCH GRANT (No. SG/PHY/P/BRM/01) from Indian Institute of Technology Guwahati, India.


\end{document}